\documentclass[conference]{IEEEtran}
\IEEEoverridecommandlockouts

\usepackage{cite}
\usepackage{amsmath,amssymb,amsfonts}
\usepackage{algorithmic}
\usepackage{graphicx}
\usepackage{textcomp}
\usepackage{xcolor}

\usepackage{multirow}
\usepackage{graphicx}
\usepackage{subcaption}
\usepackage{amsmath}
\usepackage{tabularx}
\usepackage{filecontents}
\usepackage{multirow}
\usepackage{booktabs}  
\usepackage{threeparttable}
\usepackage{colortbl}
\usepackage{siunitx}
\usepackage{tikz}
\usepackage{filecontents}

\newcommand{\Name}{Ivy}

\def\BibTeX{{\rm B\kern-.05em{\sc i\kern-.025em b}\kern-.08em
    T\kern-.1667em\lower.7ex\hbox{E}\kern-.125emX}}

\begin{document}

\title{Offline Meta-learning for Real-time Bandwidth Estimation}

\author{}
\author{\IEEEauthorblockN{Aashish Gottipati\textsuperscript{1}, Sami Khairy\textsuperscript{3}, Yasaman Hosseinkashi\textsuperscript{3}, Gabriel Mittag\textsuperscript{3}, Vishak Gopal\textsuperscript{3},\\ Francis Y. Yan\textsuperscript{2}, and Ross Cutler\textsuperscript{3}}
\IEEEauthorblockA{\textsuperscript{1}Department of Computer Science, University of Texas at Austin, Austin, Texas \\ \textsuperscript{2}Siebel School of Computing and Data Science, University of Illinois Urbana-Champaign, Urbana, Illinois \\ \textsuperscript{3}Microsoft, Redmond, Washington}}

\newcommand\copyrighttext{%
\footnotesize \textcopyright 2026 IEEE. Personal use of this material is permitted.
Permission from IEEE must be obtained for all other uses, in any current or future
media, including reprinting/republishing this material for advertising or promotional
purposes, creating new collective works, for resale or redistribution to servers or
lists, or reuse of any copyrighted component of this work in other works.}

\newcommand\copyrightnotice{%
\begin{tikzpicture}[remember picture,overlay]
\node[anchor=south,yshift=10pt] at (current page.south)
{\fbox{\parbox{\dimexpr\textwidth-\fboxsep-\fboxrule\relax}{\copyrighttext}}};
\end{tikzpicture}%
}

\maketitle
\copyrightnotice

\begin{abstract}
Real-time video applications require dynamic bitrate adjustments based on network capacity, necessitating accurate bandwidth estimation (BWE). We introduce \Name{}, a novel BWE method that leverages offline meta-learning to combat data drift and maximize user Quality of Experience (QoE). Our approach dynamically selects the most suitable BWE algorithm for current network conditions, enabling effective adaptation to changing environments without requiring live network interactions. We implemented our method in Microsoft Teams and demonstrated that \Name{} can enhance QoE by $5.9\%$ to $11.2\%$ over individual BWE algorithms and by $6.3\%$ to $11.4\%$ compared to existing online meta heuristics. Additionally, we show that our method is more data efficient compared to online meta-learning methods, achieving up to $21\%$ improvement in QoE while requiring significantly less training data.
\end{abstract}


\section{Introduction}
\label{sec:introduction}
Real-time video is ubiquitous, and an essential algorithm to maintain
user experience is bandwidth estimation (BWE). BWE refers to the process of estimating the available network capacity. Modern videoconferencing solutions continuously adjust video encoding bitrates and rely on BWE algorithms to avoid overloading the network and degrading user experience. Despite efforts in BWE~\cite{gottipati2023real, fang_reinforcement_2019, bentaleb2022bob}, there is no one size fits all solution. A recent study indicates that prior methods are unable to adapt to increasing network heterogeneity, with nearly $80\%$ of users expressing frustration and reporting degraded videoconferencing experience on popular platforms such as Zoom, Teams, and Skype~\cite{matulin2021user}.

To address the adaptability limitations of existing BWE methods, training a customized model for each network scenario has emerged as a promising solution~\cite{gottipati2023real}. This ``spatial'' customization has benefits, but network environments can still drift on the
``temporal'' dimension, due to nonstationarity (network dynamics are not constant and can vary significantly with the passage of time), leading to a degradation in BWE performance. Periodically retraining models with new data is a common strategy to address data drift, but it suffers from high retraining costs and catastrophic forgetting~\cite{mallick2022matchmaker}. When trained on new network conditions, a BWE model may abruptly discard previously learned bandwidth patterns-- effectively trading performance in new scenarios for degraded accuracy in previously well-handled network conditions. Furthermore, network environments may drift at a finer granularity, at the second level \textit{within} the same video session. We observe stark differences between production-grade bandwidth estimators when network environments drift from stable to volatile conditions. A/B testing reveals mutual regressions ranging from $1.6\%$ to $28.5\%$ in user QoE between our Quality of Service (QoS)-driven estimator and Quality of Experience (QoE)-driven BWE during live videoconferencing sessions-- with no universally optimal BWE across network conditions (see Table~\ref{table:motivate_association}).

Our key insight to address data drift is to utilize multiple custom BWEs and dynamically switch between them based on the network state. Instead of optimizing BWEs directly, we instead optimize the selection of BWEs. Choosing between BWEs offers several advantages: (1) it helps mitigate the issue of catastrophic forgetting by employing multiple specialized BWEs rather than depending on a single solution that is periodically retrained; (2) it allows us to bootstrap from prior efforts by directly reusing previously released BWEs; and (3) selection is decoupled from bandwidth estimation and can occur at longer, less frequent intervals. By operating above the estimator at a meta layer, we can achieve more accurate assessments of the network environment and user QoE.

However, selecting the optimal BWE presents three key challenges: First, QoS and QoE have an ambiguous relationship, e.g.,\ a $50\%$ throughput reduction from $4$ Mbps to $2$ Mbps may not impact user experience due to modern enhancements such as super resolution~\cite{chen2022rl}. Second, network environments fluctuate rapidly, with different BWEs reacting uniquely to transient conditions, complicating fair algorithm comparison~\cite{zhang2023bridging}. Third, practicality concerns arise as online learning approaches require thousands of live calls~\cite{gottipati2023real} with active exploration, potentially degrading QoE by increasing video freeze rates up to $79\%$~\cite{agarwal2024tarzan}, making online learning from scratch impractical for live production environments where call quality is paramount.

\begin{table*}[!t]
\vspace{0.051in}
\centering
\caption{Video QoE (Mean Opinion Score) over $400$ calls comparing QoS-driven (UKF~\cite{fang_reinforcement_2019}) and QoE-driven (R3Net~\cite{fang_reinforcement_2019}) BWEs in low bandwidth (LBW) and high bandwidth (HBW) network conditions shows that neither is universally optimal. Differences are statistically significant (Welch’s t-test, $p<0.01$), since the test evaluates mean differences relative to their standard error, allowing significance even when $95\%$ confidence intervals overlap.}
\begin{tabular}{|c|c|c|c|c|c|c|c|}
\hline
& Stable LBW & Fluctuating LBW & Burst Loss LBW & Stable HBW & Fluctuating HBW & Burst Loss HBW \\
\hline
UKF  & $\mathbf{1.20 \pm 0.02}$  & $\mathbf{2.08 \pm 0.23}$ & $\mathbf{1.11 \pm 0.03}$ & $4.17 \pm 0.05$ & $2.17 \pm 0.27$ & $1.91 \pm 0.28$ \\

R3Net  & $1.17 \pm 0.08$ & $2.00 \pm 0.04$ & $1.03 \pm 0.02$ & $\mathbf{4.24 \pm 0.01}$ & $\mathbf{2.79 \pm 0.06}$ & $\mathbf{2.44 \pm 0.12}$ \\
\hline
\hline
$\Delta$ (\%) & $2.50\%$ & $4.00\%$ & $7.00\%$ & $1.60\%$ & $28.50\%$ & $27.70\%$ \\
\hline
\end{tabular}
\label{table:motivate_association}
\vspace{-5mm}
\end{table*}

To address the previously mentioned challenges, we introduce \Name{}\footnote{A climbing plant known for its ability to thrive in various environments.}. \Name{} is trained entirely from offline telemetry logs to maximize user QoE based on observed network observations, bridging the gap between QoS and QoE. We leverage Implicit Q-learning (IQL), an offline reinforcement learning (RL) algorithm~\cite{iql}, which allows \Name{} to learn from past experiences without the need for live network interactions. Lastly, we design our method to operate at a coarse $6$-second decision interval, enabling \Name{} to more effectively monitor the impact of a given BWE on user QoE. In essence, \Name{} functions as a metapolicy that dynamically selects the most suitable BWE algorithm based on the observed network conditions, with the goal of maximizing user QoE. 
We deploy \Name{} in Microsoft Teams and utilize a diverse set of traces from fluctuating, burst loss, and stable bandwidth scenarios and nonstationary environments such as Broadband (BB), 5G, and low earth orbit (LEO) networks to analyze the performance of \Name{}. 

Thus, we make the following contributions:
\begin{enumerate}
   \item We propose \Name{}, a metapolicy that dynamically selects the most suitable BWE algorithm based on the current network conditions.
    \item Our live videoconferencing evaluations demonstrate that our offline meta-learning approach outperforms online QoS meta-heuristics, improving video QoE by $6.3\%$ to $11.4\%$.
    \item We establish the practical advantages of offline learning, with \Name{} delivering up to $21\%$ better performance than online meta-learning while requiring substantially less training data.
\end{enumerate}

\section{Related Work}
\label{sec:related}
{\bf Hybrid Systems.} Previous methods have combined learning-based agents with heuristic-based components to act as guard rails for unstable network conditions~\cite{zhang2020onrl}. Unlike these approaches that use heuristics as safeguards, \Name{} directly matches each policy with its optimal environment. While Orca~\cite{abbasloo2020classic} combines classical and RL methods for separate congestion control tasks, \Name{} coordinates the use of multiple policies for a single task: bandwidth estimation.

{\bf Exploration-Exploitation Heuristics.} Online heuristics employ exploration-exploitation cycles to greedily select control algorithms based on network feedback~\cite{du2021unified}. However, these approaches suffer from two key limitations: they prioritize short-term gains over strategic planning, and they optimize for QoS metrics (see Table~\ref{tab:qos_formulations} for common formulations) rather than QoE. As demonstrated in Table~\ref{table:motivate_association}, improvements in QoS do not consistently translate to enhanced user experience. In contrast, \Name{} addresses these limitations by directly optimizing for long-horizon user QoE through offline RL.

{\bf RL-based Metapolicies.} Several meta-approaches leverage RL to dynamically select between different control algorithms. TCP-RL~\cite{nie2019dynamic} employs online RL to choose network control algorithms by maximizing QoS metrics. Most similar to our work, Floo~\cite{zhang2023bridging} selects between congestion control algorithms using online RL to maximize QoE. While these approaches demonstrate the potential of metapolicies, they require thousands of live calls~\cite{gottipati2023real}, limiting their practicality. In contrast, \Name{} optimizes QoE from offline telemetry logs without requiring live network interactions, reducing deployment overhead in production environments. To our knowledge, \Name{} is the first method to apply offline meta-learning to BWE.
\section{Methods}
\label{sec:methods}

\subsection{Design Decisions}
Despite numerous BWEs, no universal solution exists. Recent approaches address data drift through periodic retraining \cite{gottipati2023real}; however, despite frequent retraining, our initial experiments (see Table~\ref{table:motivate_association}) suggest the potential for extreme drops in performance. Rather than relying solely on a single model, we propose dynamically selecting BWEs based on network conditions. Our design addresses three key challenges:

First, we associate BWEs with their appropriate network environments. Following prior work~\cite{zhang2023bridging}, we adopt 6-second monitoring windows~\cite{zhang2023bridging} to better estimate the QoE impact of a specific BWE in a given network setting. Additionally, we maintain the decision history and design a state space combining immediate transport layer conditions with historical performance to capture long-term decision impacts. Second, we address practical constraints of production training by leveraging offline RL, which handles the sequential nature of BWE selection while enabling long-term QoE maximization from fixed offline datasets without requiring live network interactions.
Third, we bridge the gap between QoS and user QoE. We achieve this by adopting standard network QoS signals as inputs, while optimizing directly for video QoE rather than network QoS metrics. We now introduce the key implementation details of \Name{}.

\subsection{Problem Formulation}
\label{sec:methods-2}
Given a fixed set of offline telemetry logs, we seek to learn a metapolicy that selects the most suitable bandwidth estimation policy to utilize for a given network environment.

{\bf Implicit Q-Learning.} RL problems are formulated in the context of a Markov decision process (MDP) $(\mathcal{S}, \mathcal{A}, p_0(s), p(s'|s,a),$ $ r(s,a), \gamma)$, where $\mathcal{S}$ is the state space, $\mathcal{A}$ is the action space, $p_0$ is the distribution of initial states, $p(s'|s,a)$ is the transition probabilities, $r(s, a)$ is the reward, and $\gamma$ is the discount factor. Agents interact with the MDP according to policy $\pi(a|s)$. The goal is to learn policy $\pi^*$ by maximizing the cumulative discounted rewards: 

\begin{equation}
\resizebox{0.99\columnwidth}{!}{$
    \operatorname*{argmax}_\pi \mathbb{E}_\pi \Big[\sum_{t=0}^{\infty} \gamma^t r(s_t, a_t)|s_0 \sim p_0(\cdot), a_t \sim \pi(\cdot|s_t), s_{t+1} \sim p(\cdot|s_t, a_t)\Big]
$}
\end{equation}

In contrast to online methods, offline methods are unable to query the environment and are allotted only a fixed set of offline experience. Offline learning relies on the notion of ``stitching'' 
where multiple policies can be combined into a single, more effective policy. Recent offline RL methods perform stitching by minimizing the temporal difference (TD) error in an effort to approximate the $Q$ function. Note that $Q$ captures the expected cumulative discounted rewards given state $s_t$ and action $a_t$. Given that no environment is available to query, recent offline approaches handle counterfactual queries (i.e., ``what if'' scenarios) by conservatively estimating the expected rewards~\cite{kumar2020conservative}. Improperly handling counterfactual queries can lead to poor generalization as the agent follows suboptimal trajectories based on inaccurate reward estimates for out-of-distribution actions~\cite{kumar2020conservative}. However, IQL extracts the $Q$ function without ever querying actions beyond the offline dataset. More formally, IQL~\cite{iql} formulates the TD error as follows:

\begin{figure}[t!]
    \centering
    \includegraphics[width=0.9\columnwidth, trim={6cm, 22cm, 5cm, 3cm}, clip]{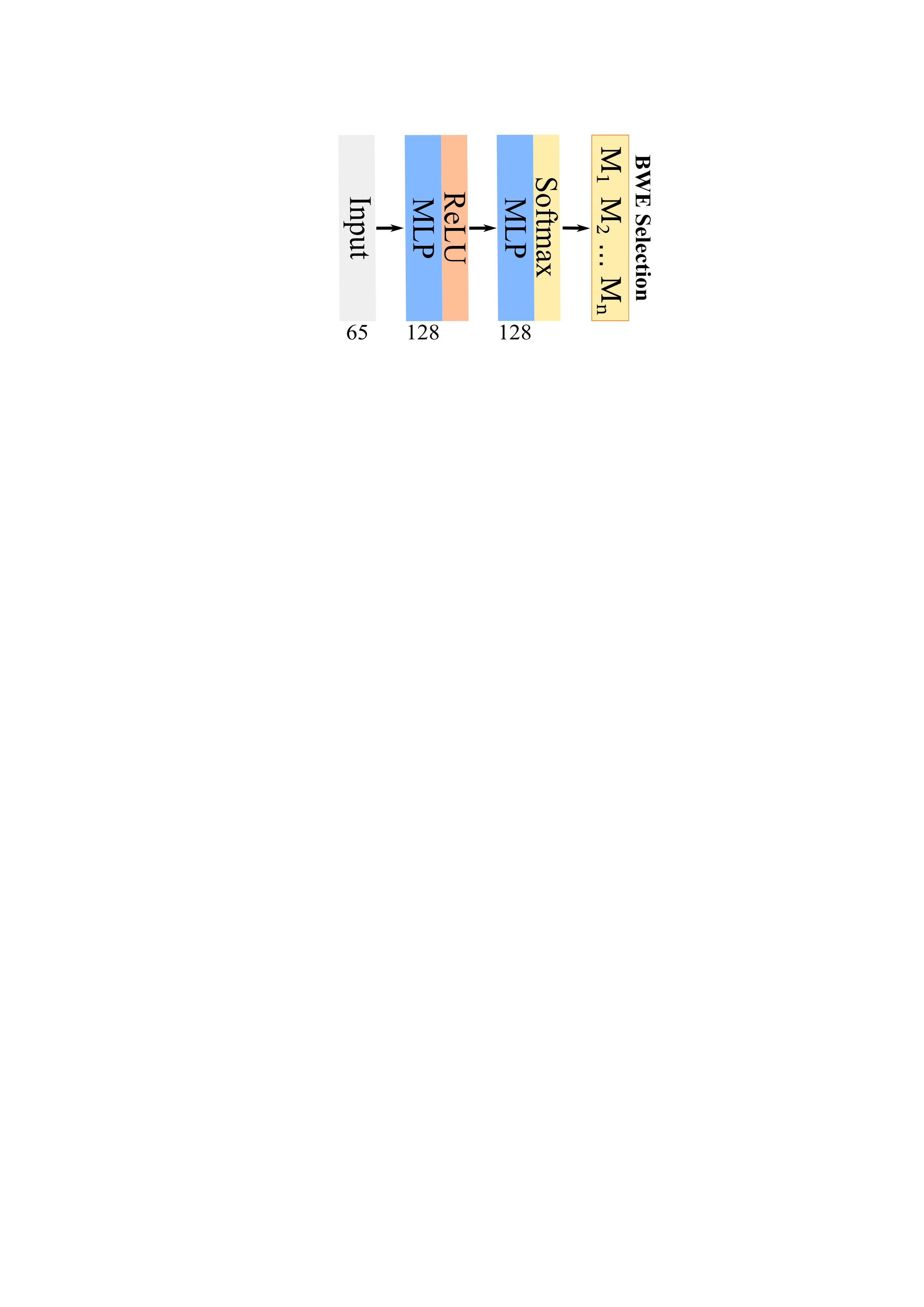}
    \caption{\Name{} Network Architecture.}
    \label{fig:architecture}
    \vspace{-6mm}
\end{figure}

\begin{equation}
\resizebox{0.99\columnwidth}{!}{$
    L(\theta) = \mathbb{E}_{(s, a, s')\sim D}\Big[(r(s,a) + \gamma \max_{\substack{a' \in \mathcal{A} \\ \text{s.t.} \pi_\beta(a'|s')>0}} Q_{\hat{\theta}}(s', a') - Q_{\theta}(s, a))^2 \Big]
$}
\end{equation}

where $D$ is the dataset, $Q_{\theta}$ is a parameterized Q-function, $Q_{\hat{\theta}}$ is a target network, and the policy is defined by $\pi(s) = \operatorname*{argmax}_a Q_{\theta}(s, a)$. IQL leverages expectile regression to approximate the state conditioned targets in the maximization step, avoiding counterfactual queries. After estimating $Q$, the policy can be extracted through advantage weighted regression~\cite{peters2007reinforcement}. By constraining the learning procedure, IQL avoids issues with out-of-distribution actions (e.g,.\ overestimating the expected reward) which can lead to poor generalization from the offline dataset.

{\bf RL Formulation.} At each decision interval $t$, we construct state $s_t = (\vec{d_t}, \vec{j_t}, \vec{l_t}, \vec{u_t}, \vec{v_t}, \vec{x_t}, \vec{z_t})$ to capture the network conditions and the previous policy decisions. Building on prior work~\cite{fang_reinforcement_2019}, state $s_t$ captures the following normalized QoS signals observed at interval $t$: one-way delay $\vec{d_t}$, packet interarrival time $\vec{j_t}$, number of lost packets $\vec{l_t}$, proportion of audio packets received $\vec{u_t}$, proportion of video packets $\vec{v_t}$, and receiving rate $\vec{x_t}$. These QoS state variables consist of multiple measurements which correspond to a coarse view of the network environment over the previous $6$ seconds, a monitoring window suggested by prior works~\cite{zhang2023bridging}. Providing the full $6$ seconds of QoS signals at $60$ ms intervals increases the state size by an order of magnitude. To strike a balance between state size and network state granularity, we opt for a more coarse grain approach and collect $10$ measurements, where each sample measurement for each QoS metric represents the average for a $600$ ms window. In contrast to the QoS signals, $\vec{z_t}$ captures the previous $5$ metapolicy actions, which enables the policy to capture the longterm impacts of prior policy decisions. We input the $65$ dimensional state $s_t$ into our actor network.

Our parameterized policy $\pi_\theta$ is represented with a two-layer multi-layer perceptron (MLP) in Figure~\ref{fig:architecture}. Given $s_t$, our policy network selects the BWE $M_i$ to run for the next time interval $[t, t+1)$ with output index $\operatorname*{argmax}_i a_t$ where $a_t \in \{0,1\}^{|M|}$ and $|M|$ is equivalent to the number of BWEs. During training, we provide QoE reward $r_t$ by averaging the observed video Mean Opinion Score (MOS) over the interval $[t, t+1)$. MOS ranges from 1 (poor) to 5 (exceptional). We employ a LSTM-based vision model that takes into account temporal distortions to compute the video MOS per received video frame. Estimates were shown to have $99\%$ correlation with user visual experience~\cite{mittag2023lstm}. While prior work defines QoE using QoS metrics (throughput, freezes, and loss rate), we choose MOS as it is the ``industry standard'' for user experience evaluation (conferencing software commonly requests 5-star ratings post-call).

\subsection{Data Collection}
\label{sec:methods-3}
Offline learning is bound by the best combination of policies present in the fixed offline dataset. To ensure adequate coverage over the space of metapolicies, we collected logs from $1000$ videoconferencing calls using a random policy. Videoconferencing calls were conducted for $2$ minutes over emulated links where network traces were sampled from diverse scenarios, encompassing fluctuating bandwidth, burst loss patterns, and stable connections with bandwidths ranging from $0.1$ Mbps to $8$ Mbps, as well as real-world nonstationary environments such as BB, 5G, and LEO. To enhance dataset diversity, we uniformly sampled environmental conditions, with packet loss rates ranging from $0$ to $0.25$ and round-trip delay times varying from $40$ to $60$ ms.

At each decision interval, the random policy uniformly selects among a pool of BWEs (e.g.,\ UKF and R3Net) to run for the next interval. Our telemetry logs the relevant trajectory information: the history of metapolicy actions, the QoS network metrics, and the average estimated video MOS. Experience was collected once prior to training.

\subsection{Training Details and Implementation}
We represent our metapolicy with a two-layer neural network (see Figure~\ref{fig:architecture}). Each layer consists of $128$ neurons. We utilize a ReLU activation function for the first hidden layer and apply a softmax to the output layer. Although prior works leverage recurrent models for BWE~\cite{fang_reinforcement_2019}, we observed no QoE improvements from adopting a more complex model architecture. We hypothesize that the $65$-dimensional state vector already captures the necessary features for effective BWE selection.  \Name{} was trained on offline experience collected according to the procedure detailed in Section~\ref{sec:methods-3}. Additionally, we experimented with decision intervals of $1.2$s, $3.0$s, and $4.8$s (multiples of $60$ms); however, we observed no statistically significant difference in user QoE when compared to $6.0$s decision intervals. Each videoconferencing call was treated as a single training sample. The batch size was set to $128$ and default IQL hyperparameters were utilized. \Name{} was trained for $100$ epochs without a single live network interaction.

We implement and train \Name{} in Pytorch. \Name{}
is deployed directly into our RTC media stack for receiver-side bandwidth estimation. \Name{}'s lightweight architecture introduces minimal resource overhead: the memory footprint increases by $0.2$ MB, while maintaining real-time bandwidth estimates on Microsoft Teams compatible devices at the industry-standard $60$ ms granularity.
\section{Evaluation}
\label{sec:eval}

The key takeaways from our evaluations are as follows:
\begin{enumerate}
\item \Name{} enhances video QoE by $5.9\%$ to $11.2\%$ over individual BWE estimators.
\item Our offline QoE-driven approach outperforms online QoS meta-heuristics, delivering $6.3\%$ to $11.4\%$ video QoE improvements.
\item \Name{} effectively handles network nonstationarity, achieving up to $6\%$ video QoE improvements across diverse environments (5G, BB, and LEO) compared to baseline estimators.
\item Our offline meta-learning methodology delivers up to $28\%$ better performance compared to online meta-learning while using the same amount of training data.
\end{enumerate}

\subsection{Setup}
{\bf Environment Setup.} We evaluate \Name{} on two complementary testbeds with A/B testing. We conduct peer-to-peer videoconferencing calls with BWEs running at $60$ms granularity and metapolicy decisions executing every $6$ seconds. In our primary setup, we utilize a Linux-based network router with netem to emulate various network conditions across randomly selected machine pairs within the same datacenter. Tests run on holdout traces covering fluctuating, burst loss, and stable conditions in both low bandwidth (LBW) and high bandwidth (HBW) environments. For additional validation, we leverage Cloudlab~\cite{duplyakin2019design}, AlphaRTC~\cite{eo2022opennetlab} (an open-source RTC stack), and the Mahimahi network emulation tool~\cite{netravali2015mahimahi} and evaluate performance on public network traces~\cite{he2024designing} from real-world 5G, BB, and LEO satellite environments.

\begin{figure}[t!]
    \centering
    \includegraphics[width=0.9\columnwidth, trim={0cm, 0cm, 0cm, 0cm}, clip]{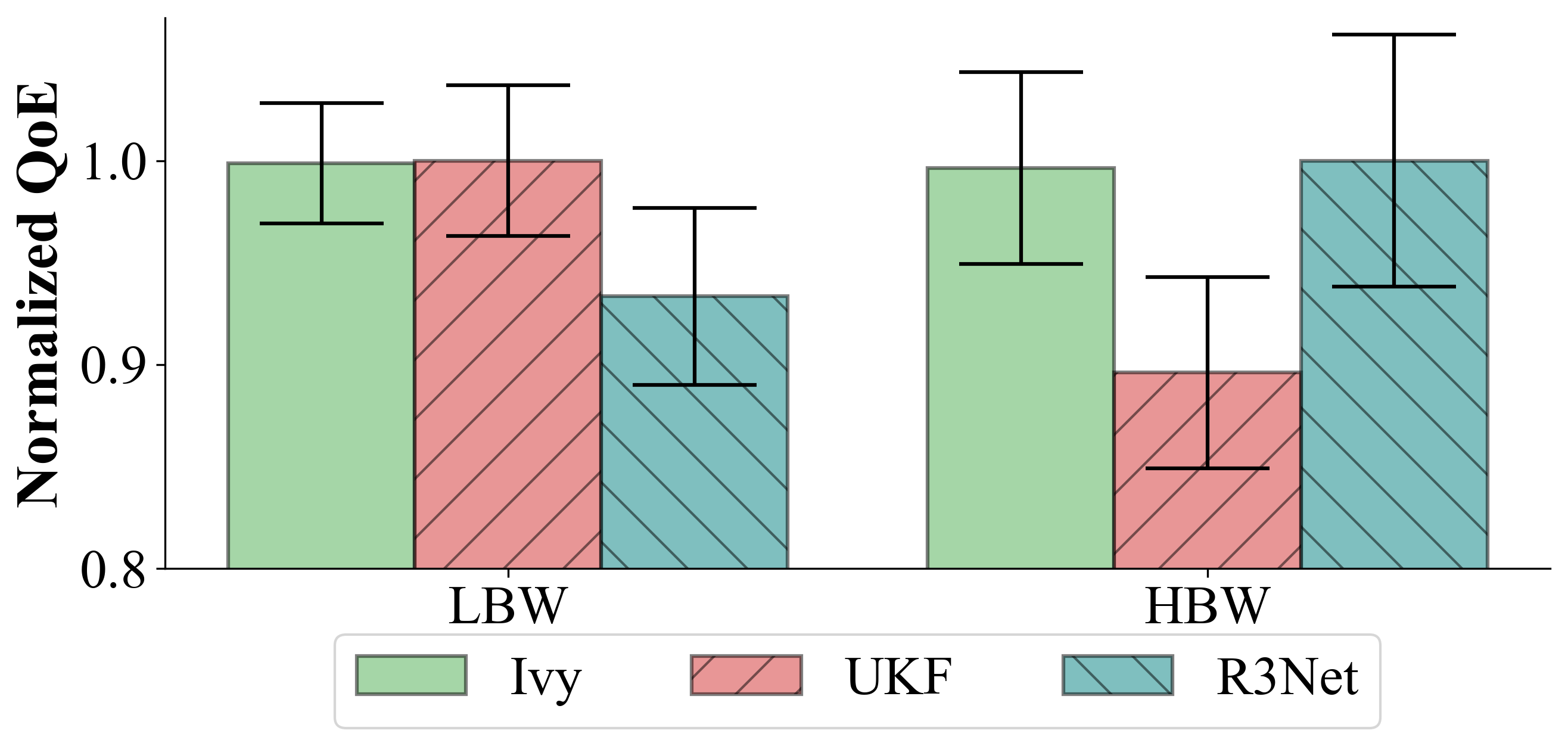}
    \caption{Comparison between \Name{} and Individual Estimators.}
    \label{fig:individual_comp}
    \vspace{-5mm}
\end{figure}

{\bf Baselines.} We evaluate \Name{} against three baseline categories. (1) We compare against the following individual estimators. UKF~\cite{fang_reinforcement_2019}, an unscented Kalman filter optimized for QoS; R3Net~\cite{fang_reinforcement_2019}, an LSTM-based estimator trained via offline RL to maximize video MOS that has previously shown a mix of improvements and regressions across network environments in comparison to production-grade UKF. (2) We compare against online exploration-exploitation and rule-based metapolicies. For exploration-exploitation, we implement $6$-second decision intervals. We follow prior approaches~\cite{du2021unified} and begin exploration with the more conservative QoS-based BWE UKF. The estimator achieving higher QoS utility during exploration (see Table~\ref{tab:qos_formulations} for QoS formulations) is selected for the remainder of the call. Additionally, we compare against three rule-based approaches: jitter-based, delta-based, and a combination of the two. The jitter threshold is set to $25$ ms. In contrast, for the delta-based rule, if the difference between UKF and R3Net exceeds one standard deviation from prior estimates, then the policy defaults to the QoS-driven UKF estimator with R3Net running during stable conditions. Rule-based parameters were selected based on empirical observations.
(3) We baseline against online RL meta-models trained via Proximal Policy Optimization (PPO)~\cite{schulman2017proximal}, a state-of-the-art online RL algorithm. We train multiple online meta-models on live videoconferencing calls from traces drawn from the same training trace distribution as \Name{}.

{\bf Metrics.} 
We define QoE in terms of the estimated video MOS. We normalize the reported MOS values by the best performing BWE in each network setting. Unless otherwise specified, we report the normalized mean video MOS with two-sided $95$\% confidence intervals. 

\subsection{Comparison with Baseline Estimators}
\label{sec:eval-1}
{\bf Remark: \Name{} intelligently adapts to network conditions by selecting the most suitable bandwidth estimator for the current environment.} Specifically, in LBW environments, \Name{} performs on par with UKF while improving video QoE by $6.2\%$ over R3Net (see Figure~\ref{fig:individual_comp}). In HBW environments, \Name{} matches R3Net's performance while delivering an $11.2\%$ QoE improvement over UKF. While the aggregated improvements may appear modest on a normalized scale, their practical significance is substantial. User experience in real-time video is highly non-linear; users are far more sensitive to sudden, jarring events such as freezes than to minor fluctuations in quality. \Name{}'s primary value lies not in small, constant gains but in its ability to mitigate catastrophic failures. For instance, \Name{} prevents the sharp, perceptible quality drops that frustrate users by avoiding situations where a single estimator's performance can plummet by $27.7\%$ under unfavorable network conditions. In practical terms, \Name{} prevents the user experience from degrading from Poor (MOS $2.44$) to Bad (MOS $1.91$) during burst loss events (Table~\ref{table:motivate_association}). By consistently selecting the better-performing algorithm, \Name{} delivers a more reliable and stable experience, where the key benefit is the absence of severe degradation.

\subsection{Offline Learning vs. Online Meta Heuristics}
\label{sec:eval-2}

\begin{table}[t!]
\vspace{0.051in}
\centering
\caption{QoS Objectives Tested.}
\begin{tabular}{ccc}
\toprule
\textbf{Objective} & \textbf{Formulation} \\
\midrule
Vivace & $\text{rate}^t - \beta \cdot \text{rate} \cdot \frac{d\text{RTT}}{dT} - \gamma \cdot \text{loss} \cdot \text{rate}$                     \\
Power Variant   & $\text{rate} \cdot (1-\text{loss})$ / \text{delay}                   \\
Power   & \text{rate} / \text{delay}               \\
Throughput   & $\text{rate}$\\
\bottomrule
\end{tabular}
\label{tab:qos_formulations}
\vspace{-6mm}
\end{table}

\begin{figure*}[t!]
    \centering
    \begin{subfigure}[t]{0.33\textwidth}
        \centering
        \includegraphics[width=\textwidth]{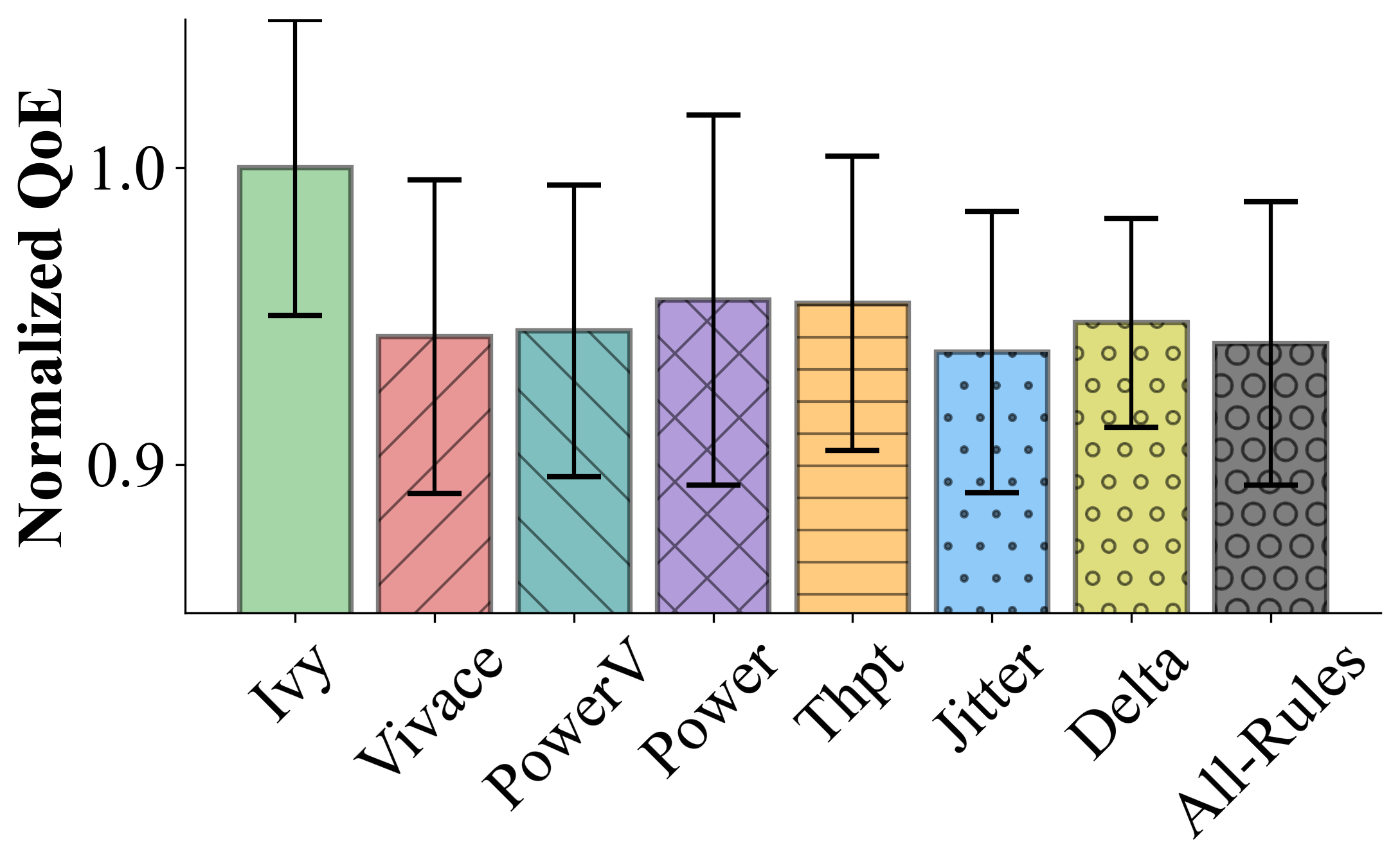}
        \caption{All Environments.}
    \end{subfigure}%
    ~     
        \begin{subfigure}[t]{0.33\textwidth}
        \centering
        \includegraphics[width=\textwidth]{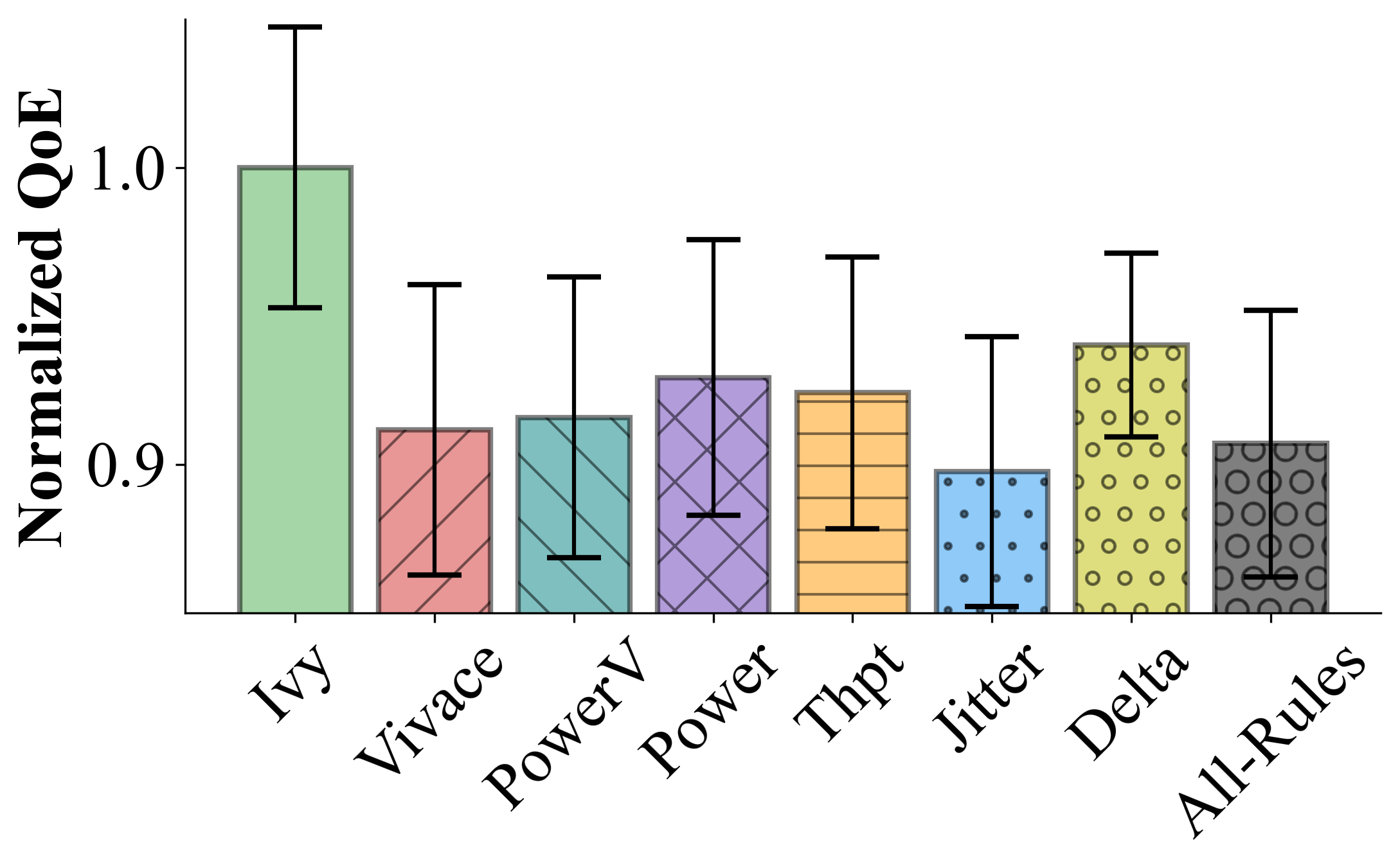}
        \caption{High Bandwidth.}
    \end{subfigure}%
    ~ 
    \begin{subfigure}[t]{0.33\textwidth}
        \centering
        \includegraphics[width=\textwidth]{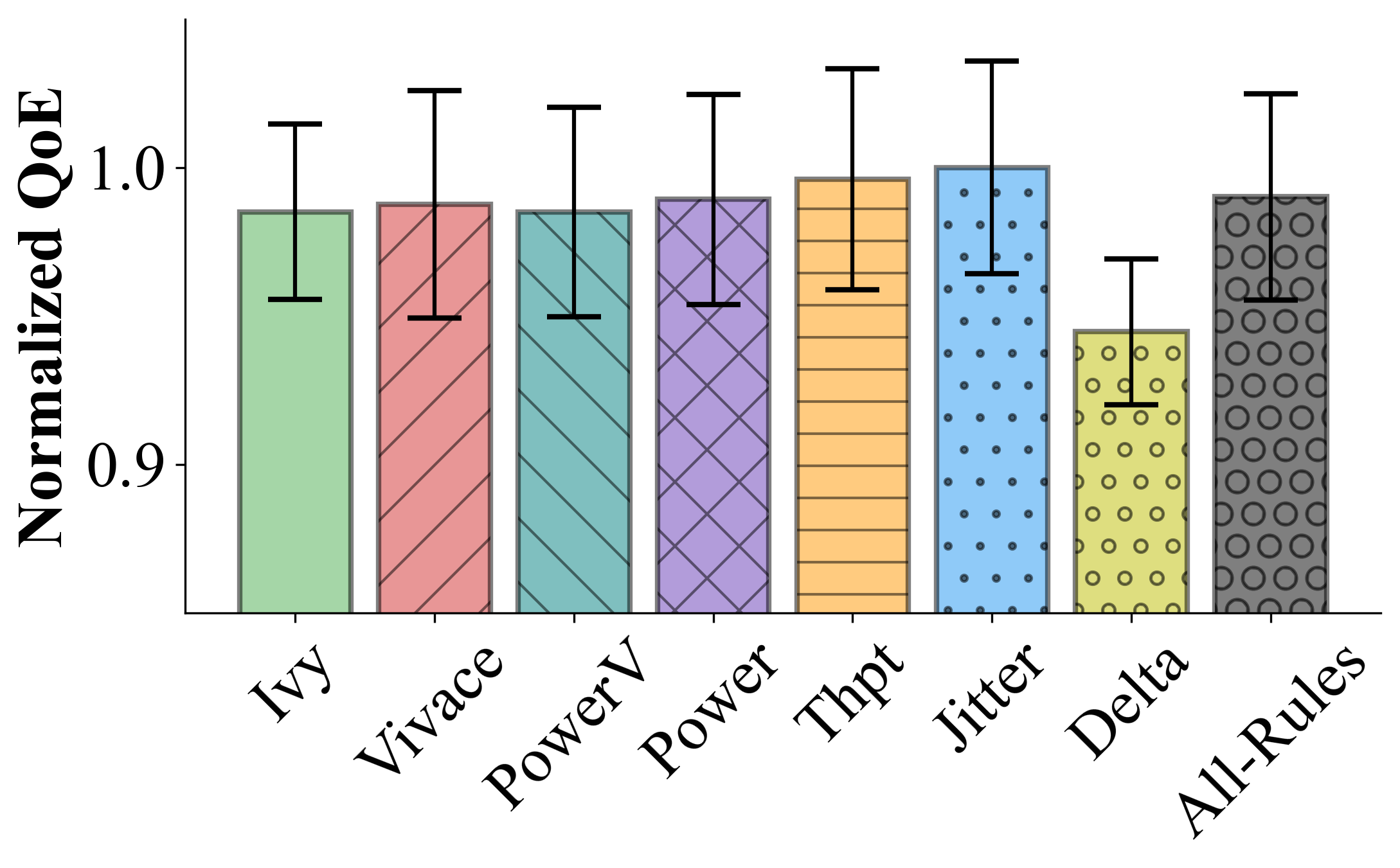}
        \caption{Low Bandwidth.}
    \end{subfigure}%
    \caption{\Name{} vs. Online QoS-based Meta Heurstics.}
    \label{fig:online_heuristics}
    \vspace{-4mm}
\end{figure*}

{\bf Remark: Divergence between QoE and QoS leads to suboptimal performance in HBW environments.} After conducting $300$ videoconferencing calls per approach, we find that \Name{} improves video QoE by approximately $6.3\%$ to $11.4\%$ compared to common online QoS meta heuristics and hybrid methods (see Figure~\ref{fig:online_heuristics}). Despite being trained completely offline without network interactions and maintaining a fixed runtime policy, \Name{} achieves superior performance by directly optimizing for user experience rather than network metrics. This advantage is most pronounced in HBW environments, where QoS-driven heuristics favor UKF to optimize transport-layer metrics, while \Name{} selects R3Net, prioritizing user experience. The performance gap stems from the non-linear relationship between QoS and QoE in videoconferencing-- maximizing QoE often requires QoS trade-offs, particularly when additional throughput doesn't translate to better user experience. By optimizing for MOS instead of traditional QoS metrics, \Name{} validates our thesis that effective videoconferencing BWE requires directly targeting user experience measurements.

\subsection{Strength in Multiple Estimators}
\label{sec:eval-4}

\begin{figure}[t!]
    \centering
    \includegraphics[width=\columnwidth, trim={0cm, 0cm, 0cm, 0cm}, clip]{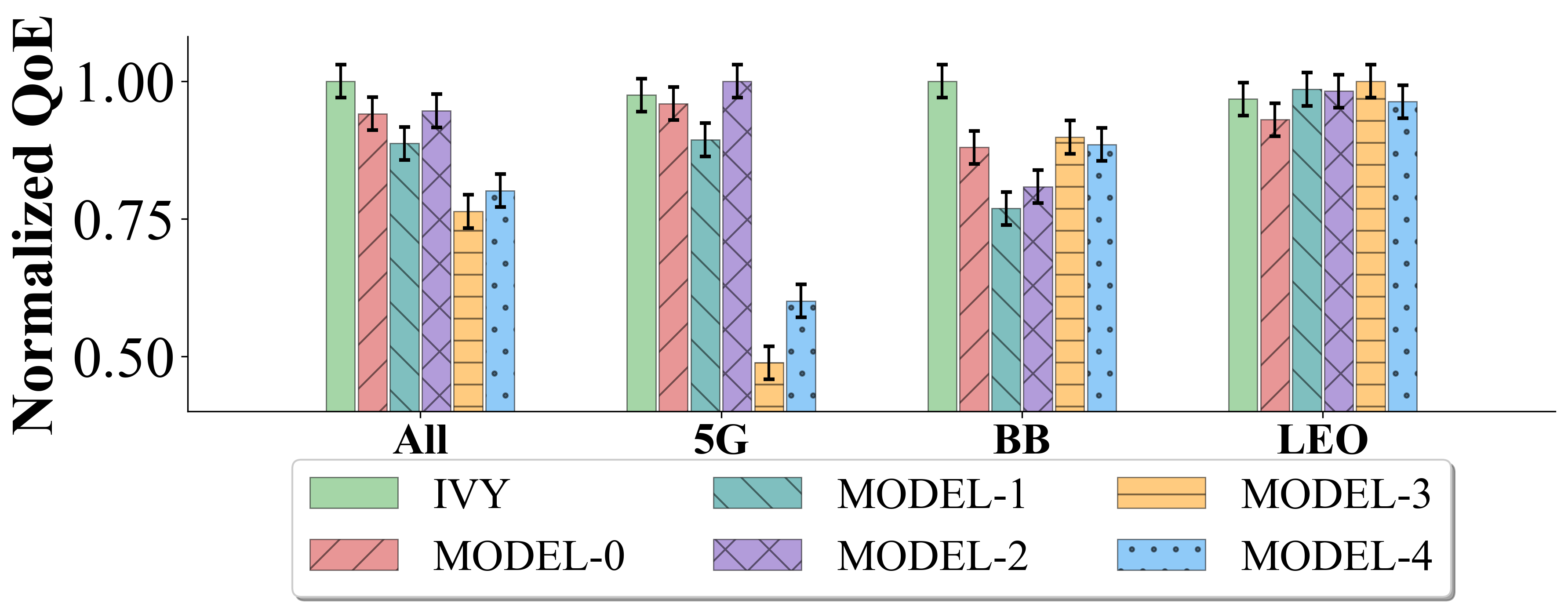}
    \caption{Comparison between \Name{} and Individual Estimators in Nonstationary Environments.}
    \label{fig:five_models}
    \vspace{-2mm}
\end{figure}

In contrast to our prior evaluations, we now evaluate \Name{} in an opensource testbed with AlphaRTC~\cite{eo2022opennetlab}. Using public 5G, BB, and LEO network traces~\cite{he2024designing} and the Mahimahi emulation tool~\cite{netravali2015mahimahi}, we collect offline telemetry from $1000$ videoconferencing calls with WebRTC's Google Congestion Control (GCC)~\cite{carlucci2016analysis}, a widely adopted rate control algorithm. Inspired by~\cite{agarwal2024tarzan}, we then train five new BWEs with offline RL, which we refer to as model-$0$ through model-$4$. Each BWE is trained on differing subsets of offline telemetry, with no one model being trained on all data. Lastly, we follow the data collection process from Section~\ref{sec:methods} and retrain \Name{}. We conduct $30$ videoconferencing calls per BWE in this new environment and report our observed QoE values in Figure~\ref{fig:five_models}.

{\bf Remark: Metalearning enables robust adaption to nonstationary network environments.} Rather than relying on a single, generalizable model, \Name{} leverages multiple specialized models. \Name{} intelligently selects the most suitable BWE to run based on the current network state, leading to a $6\%$ improvement in video QoE on average compared against our baseline estimators in these nonstationary environments. Specifically, we observe no statistically significant regressions in both LEO and 5G cases; however, we observe a $12\%$ improvement in BB. By dynamically selecting the best estimator for the current network conditions, \Name{} enhances video QoE in nonstationary network environments.

\subsection{Offline vs. Online Learning}
\label{sec:eval-5}
\begin{figure}[t!]
    \centering
    \includegraphics[width=\columnwidth, trim={0cm, 0cm, 0cm, 0cm}, clip]{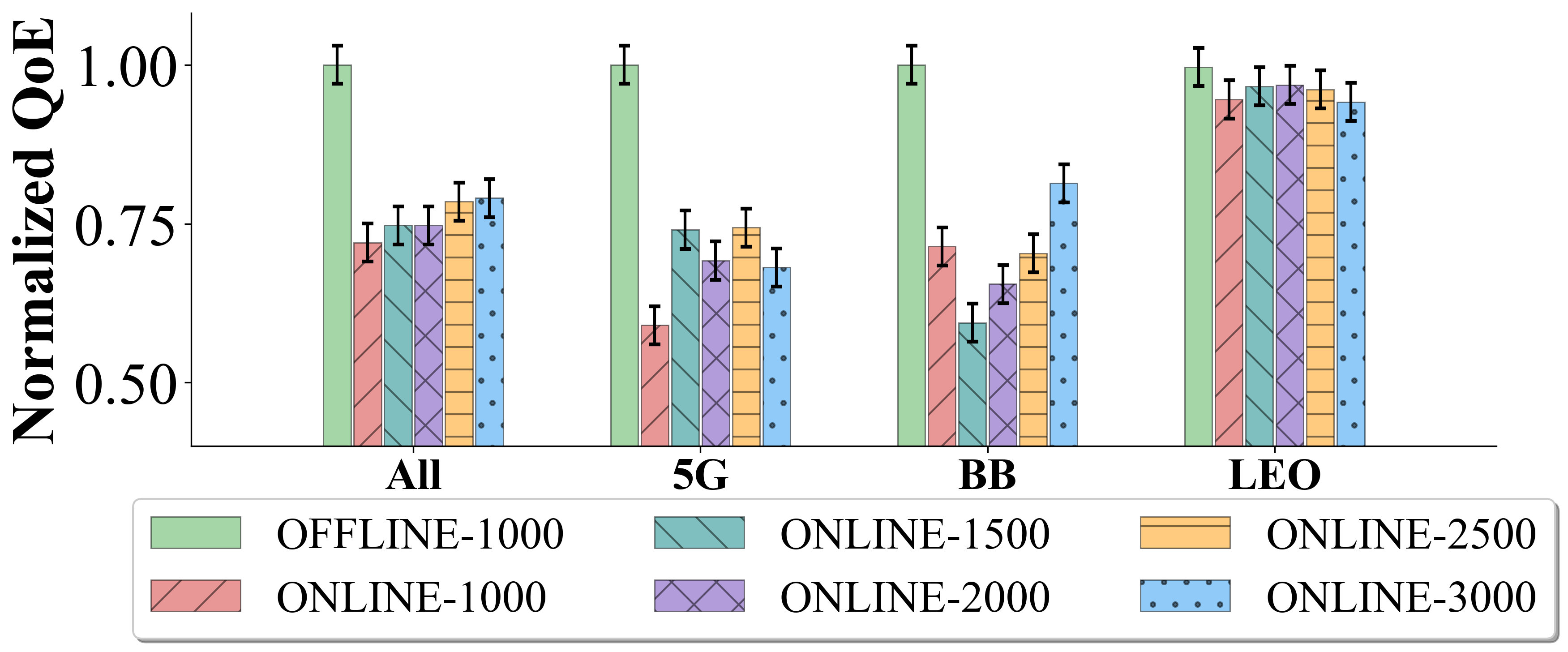}
    \caption{Offline vs Online Meta-learning.}
    \label{fig:online_v_offline}
    \vspace{-7mm}
\end{figure}

We now compare \Name{}'s offline meta-learning approach with online meta-learning. Specifically, we compare with online metapolicies trained via PPO. For this comparison, we train five online meta-models from scratch to maximize video QoE. Each policy selects among the five BWEs described in Section~\ref{sec:eval-4}. We vary the number of training calls from $1000$ to $3000$ in increments of $500$, which we denote online-$1000$ through online-$3000$. Following the same experimental setup as our previous evaluation, we present our results in Figure~\ref{fig:online_v_offline}.

{\bf Remark: Offline learning provides superior data efficiency.} For a fixed number of videoconferencing calls, our results demonstrate that \Name{}'s offline approach improves video QoE by approximately $21\%$ to $28\%$ over its online counterpart (see Figure~\ref{fig:online_v_offline}). This performance advantage is achieved despite \Name{} being trained completely offline on a limited dataset, without the need for online network interactions during training. Specifically, offline learning enhances data utilization by performing multiple passes over previously collected experience, thereby extracting more value from each observation. In contrast, online training proceeds incrementally, learning from each new experience only once. Thus, with $1000$ previously collected videoconferencing telemetry logs, \Name{}'s offline metalearning approach significantly outperforms its online counterpart trained with an equivalent number of live calls by $28\%$.
\section{Conclusion}
\label{sec:conclusion}

We position our work as advancing the robustness of ML-based systems and focus primarily on addressing data drift, with the key insight that dynamically selecting the optimal BWE algorithm for current conditions enables better adaptation to changing environments. We introduce \Name{}, a QoE-driven metapolicy that dynamically selects the most suitable BWE for given network conditions. Unlike previous approaches, \Name{} learns from offline telemetry logs without live network interaction, reducing deployment complexity and enhancing real-world practicality. Evaluations show our offline meta-learning approach improves QoE by up to $11.4\%$ over online QoS-based meta heuristics. Future work will explore a hybrid approach, using the offline-trained model as a robust starting point for selective online fine-tuning to adapt to novel network environments.
\bibliographystyle{IEEEtran}
\bibliography{IC3-AI, bibliography}

\begin{thebibliography}{10}
\providecommand{\url}[1]{#1}
\csname url@samestyle\endcsname
\providecommand{\newblock}{\relax}
\providecommand{\bibinfo}[2]{#2}
\providecommand{\BIBentrySTDinterwordspacing}{\spaceskip=0pt\relax}
\providecommand{\BIBentryALTinterwordstretchfactor}{4}
\providecommand{\BIBentryALTinterwordspacing}{\spaceskip=\fontdimen2\font plus
\BIBentryALTinterwordstretchfactor\fontdimen3\font minus \fontdimen4\font\relax}
\providecommand{\BIBforeignlanguage}[2]{{%
\expandafter\ifx\csname l@#1\endcsname\relax
\typeout{** WARNING: IEEEtran.bst: No hyphenation pattern has been}%
\typeout{** loaded for the language `#1'. Using the pattern for}%
\typeout{** the default language instead.}%
\else
\language=\csname l@#1\endcsname
\fi
#2}}
\providecommand{\BIBdecl}{\relax}
\BIBdecl

\bibitem{gottipati2023real}
A.~Gottipati, S.~Khairy, G.~Mittag, V.~Gopal, and R.~Cutler, ``Real-time bandwidth estimation from offline expert demonstrations,'' \emph{arXiv preprint arXiv:2309.13481}, 2023.

\bibitem{fang_reinforcement_2019}
\BIBentryALTinterwordspacing
J.~Fang, M.~Ellis, B.~Li, S.~Liu, Y.~Hosseinkashi, M.~Revow, A.~Sadovnikov, Z.~Liu, P.~Cheng, S.~Ashok, D.~Zhao, R.~Cutler, Y.~Lu, and J.~Gehrke, ``Reinforcement learning for bandwidth estimation and congestion control in real-time communications,'' 2019, arXiv:1912.02222 [cs]. [Online]. Available: \url{http://arxiv.org/abs/1912.02222}
\BIBentrySTDinterwordspacing

\bibitem{bentaleb2022bob}
A.~Bentaleb, M.~N. Akcay, M.~Lim, A.~C. Begen, and R.~Zimmermann, ``Bob: Bandwidth prediction for real-time communications using heuristic and reinforcement learning,'' \emph{IEEE Transactions on Multimedia}, vol.~25, pp. 6930--6945, 2022.

\bibitem{matulin2021user}
M.~Matulin, {\v{S}}.~Mrvelj, B.~Abramovi{\'c}, T.~{\v{S}}o{\v{s}}tari{\'c}, and M.~{\v{C}}ejvan, ``User quality of experience comparison between skype, microsoft teams and zoom videoconferencing tools,'' in \emph{International Conference on Future Access Enablers of Ubiquitous and Intelligent Infrastructures}.\hskip 1em plus 0.5em minus 0.4em\relax Springer, 2021, pp. 299--307.

\bibitem{mallick2022matchmaker}
A.~Mallick, K.~Hsieh, B.~Arzani, and G.~Joshi, ``Matchmaker: Data drift mitigation in machine learning for large-scale systems,'' \emph{Proceedings of Machine Learning and Systems}, vol.~4, pp. 77--94, 2022.

\bibitem{chen2022rl}
K.~Chen, H.~Wang, S.~Fang, X.~Li, M.~Ye, and H.~J. Chao, ``Rl-afec: adaptive forward error correction for real-time video communication based on reinforcement learning,'' in \emph{Proceedings of the 13th ACM Multimedia Systems Conference}, 2022, pp. 96--108.

\bibitem{zhang2023bridging}
J.~Zhang, Y.~Zhang, E.~Dong, Y.~Zhang, S.~Ren, Z.~Meng, M.~Xu, X.~Li, Z.~Hou, Z.~Yang \emph{et~al.}, ``Bridging the gap between $\{$QoE$\}$ and $\{$QoS$\}$ in congestion control: A large-scale mobile web service perspective,'' in \emph{2023 USENIX Annual Technical Conference (USENIX ATC 23)}, 2023, pp. 553--569.

\bibitem{agarwal2024tarzan}
N.~Agarwal, R.~Pan, F.~Y. Yan, and R.~Netravali, ``Tarzan: Passively-learned real-time rate control for video conferencing,'' \emph{arXiv preprint arXiv:2410.03339}, 2024.

\bibitem{iql}
I.~Kostrikov, A.~Nair, and S.~Levine, ``Offline reinforcement learning with implicit q-learning,'' \emph{arXiv preprint arXiv:2110.06169}, 2021.

\bibitem{zhang2020onrl}
H.~Zhang, A.~Zhou, J.~Lu, R.~Ma, Y.~Hu, C.~Li, X.~Zhang, H.~Ma, and X.~Chen, ``Onrl: improving mobile video telephony via online reinforcement learning,'' in \emph{Proceedings of the 26th Annual International Conference on Mobile Computing and Networking}, 2020, pp. 1--14.

\bibitem{abbasloo2020classic}
S.~Abbasloo, C.-Y. Yen, and H.~J. Chao, ``Classic meets modern: A pragmatic learning-based congestion control for the internet,'' in \emph{Proceedings of the Annual conference of the ACM Special Interest Group on Data Communication on the applications, technologies, architectures, and protocols for computer communication}, 2020, pp. 632--647.

\bibitem{du2021unified}
Z.~Du, J.~Zheng, H.~Yu, L.~Kong, and G.~Chen, ``A unified congestion control framework for diverse application preferences and network conditions,'' in \emph{Proceedings of the 17th International Conference on emerging Networking EXperiments and Technologies}, 2021, pp. 282--296.

\bibitem{nie2019dynamic}
X.~Nie, Y.~Zhao, Z.~Li, G.~Chen, K.~Sui, J.~Zhang, Z.~Ye, and D.~Pei, ``Dynamic tcp initial windows and congestion control schemes through reinforcement learning,'' \emph{IEEE Journal on Selected Areas in Communications}, vol.~37, no.~6, pp. 1231--1247, 2019.

\bibitem{kumar2020conservative}
A.~Kumar, A.~Zhou, G.~Tucker, and S.~Levine, ``Conservative q-learning for offline reinforcement learning,'' \emph{Advances in Neural Information Processing Systems}, vol.~33, pp. 1179--1191, 2020.

\bibitem{peters2007reinforcement}
J.~Peters and S.~Schaal, ``Reinforcement learning by reward-weighted regression for operational space control,'' in \emph{Proceedings of the 24th international conference on Machine learning}, 2007, pp. 745--750.

\bibitem{mittag2023lstm}
G.~Mittag, B.~Naderi, V.~Gopal, and R.~Cutler, ``Lstm-based video quality prediction accounting for temporal distortions in videoconferencing calls,'' in \emph{ICASSP 2023-2023 IEEE International Conference on Acoustics, Speech and Signal Processing (ICASSP)}.\hskip 1em plus 0.5em minus 0.4em\relax IEEE, 2023, pp. 1--5.

\bibitem{duplyakin2019design}
D.~Duplyakin, R.~Ricci, A.~Maricq, G.~Wong, J.~Duerig, E.~Eide, L.~Stoller, M.~Hibler, D.~Johnson, K.~Webb \emph{et~al.}, ``The design and operation of $\{$CloudLab$\}$,'' in \emph{2019 USENIX annual technical conference (USENIX ATC 19)}, 2019, pp. 1--14.

\bibitem{eo2022opennetlab}
J.~Eo, Z.~Niu, W.~Cheng, F.~Y. Yan, R.~Gao, J.~Kardhashi, S.~Inglis, M.~Revow, B.-G. Chun, P.~Cheng \emph{et~al.}, ``Opennetlab: Open platform for rl-based congestion control for real-time communications,'' in \emph{Proceedings of the 6th Asia-Pacific Workshop on Networking}, 2022, pp. 70--75.

\bibitem{netravali2015mahimahi}
R.~Netravali, A.~Sivaraman, S.~Das, A.~Goyal, K.~Winstein, J.~Mickens, and H.~Balakrishnan, ``Mahimahi: accurate $\{$Record-and-Replay$\}$ for $\{$HTTP$\}$,'' in \emph{2015 USENIX Annual Technical Conference (USENIX ATC 15)}, 2015, pp. 417--429.

\bibitem{he2024designing}
Z.~He, A.~Gottipati, L.~Qiu, X.~Luo, K.~Xu, Y.~Yang, and F.~Y. Yan, ``Designing network algorithms via large language models,'' in \emph{Proceedings of the 23rd ACM Workshop on Hot Topics in Networks}, 2024, pp. 205--212.

\bibitem{schulman2017proximal}
J.~Schulman, F.~Wolski, P.~Dhariwal, A.~Radford, and O.~Klimov, ``Proximal policy optimization algorithms,'' \emph{arXiv preprint arXiv:1707.06347}, 2017.

\bibitem{carlucci2016analysis}
G.~Carlucci, L.~De~Cicco, S.~Holmer, and S.~Mascolo, ``Analysis and design of the google congestion control for web real-time communication (webrtc),'' in \emph{Proceedings of the 7th International Conference on Multimedia Systems}, 2016, pp. 1--12.

\end{thebibliography}
\end{document}